# Generativism: the new hybrid

Author B. Mairéad Pratschke (University of Manchester)
Mairead.Pratschke@manchester.ac.uk

Abstract

Generative Artificial Intelligence (GenAI) in Education has in a few short months moved from being the topic of discussion around speculative education futures to a very concrete reality. It is clear that the future of education, as all industries, is collaboration with GenAI. GenAI's attributes make it well suited for social and constructivist approaches to learning that value collaboration, community and the construction of knowledge and skills through active learning. This article presents an approach to designing education in collaboration with GenAI, based on digital education frameworks adapted for this new hybrid of the AI age.

Introduction

We have entered the era of Web 4.0, the so-called symbiotic web, characterised by human-computer integration and collaboration. Web 3.0 and 4.0 technologies are driving the Sixth Wave of Innovation and the 4th Industrial Revolution. No new technology since the introduction of the World Wide Web has been more disruptive than generative AI, which exploded onto the scene in late 2022, forcing us to consider the future of education.

Two or so decades ago, a similar shift began, when the transition from the original read-write web (later known as Web 1.0) to a more user-friendly, social web (Web 2.0) allowed users to generate content, share and collaborate. For online learning, this shift was transformational, as students could now interact with each other, work together, and share and reflect on ideas – just as they would in a physical classroom. Changes facilitated by Web 2.0 technologies took place in traditional classrooms as well. Campus instructors "flipped" their classes, using face-to-face time to delve deeper into topics, while assigning what used to be lecture content for students to cover in their own time. These approaches, which rethought the normal hierarchy of classroom activities, were part of a wider shift from the instructor-centred "sage on the stage" model of teaching to the more student-centred "guide on the side" approach, in which the student is an active participant and a co-creator of their own learning.

A new learning theory emerged early in this period, which reflected both the increasingly important role of technology in education and the shift to student as an constructor of their own learning. This theory, Connectivism (2005), saw students as nodes in the digital network and active agents in their own learning. This idea of the digital network as the place where knowledge resides, and to which students have access as agents in their own learning, is an important link between the digital world of Web 2.0 and the generative world of Web 4.0. And this shift, from instructor-led learning in a physical classroom, to student-centred learning in the digital ecosystem, laid the groundwork for Education 4.0 and what lies ahead.

Education 4.0 is a model of education that uses digital and AI tools to offer a personalised and accessible educational experience, based on active pedagogies and assessed using a variety of project-, challenge- and competency-based approaches. The instructor still matters but their role has changed from disseminator of knowledge to curator, guide, facilitator, and potentially, data analyst. Learning spaces are no longer defined by the simple binary of online or on-campus, but instead in smart, connected spaces, facilitated by the digital and now growing AI ecosystem of tools.

This integration of human + computer capabilities enabled by Web 4.0 tools clearly requires a new approach to the design of learning experiences. That approach is called Generativism.

Generativism

Generativism (n.) describes the symbiotic approach to teaching and learning with generative AI (GenAI). It is grounded in the principle of learning as a process, and employs constructivist and collaborative approaches to instruction. Codesign with GenAI is the defining feature of generativism as a practice, whereby knowledge is generated in collaboration with GenAI, through learning activities that are codesigned with, facilitated by, and assessed with GenAI. This approach requires a shift in assessment practices, from assessing learning as output to assessing learning as a process (also known as assessment as learning), because assessment is part of the learning process and does not exist in isolation from it.

Generativism in practice can be summarised as:
1) codesign of the learning experience in collaboration with GenAI;
2) codelivery of the learning activities and assessments in collaboration with GenAI;
3) assessment of learning as a process in collaboration with GenAI.

This approach builds on learning theories and approaches, including:
- Social learning (Bandura, 1977), which stresses the importance of collaboration;
- Constructivism (Piaget, Vygotsky, von Glaserfeld, 1970s), which highlights the need for active learning in order for students to construct meaning;
- Experiential learning (Kolb, 1984), which highlights the need for reflective observation and active experimentation as part of the process; and
- Connectivism (Siemens, Downes, 2005), which views student as nodes in a digital network and in which they are active agents in their own learning.

GenAI attributes

In order to collaborate effectively with GenAI, it is helpful to understand how the machine works and what it is capable of. GenAI tools are built on Large Language Models (LLMs) designed to both process and understand human language. These natural language understanding (NLU) and natural language processing (NLP) capabilities are the key to thinking about how and where we might best use GenAI tools in education. Chatbots provide a conversational interface for learners to interact with, where NLU and NLP is used to analyse and understand language in social interactions, to answer questions, to provide feedback and to engage in dialogue. The best known are OpenAI's ChatGPT, Google's Bard, Microsoft's Bing Chat, Perplexity and Anthropic's Claude. There are also an array of AI tutors and buddies, such as the Khan Academy's Khanmigo and Snapchat's AI buddy, as well as Pi, Poe, and many more. These chatbots, marketed as buddies and tutors, are natural partners in social learning.

The release of ChatGPT4 in March 2023 signalled a rapid acceleration in the capabilities of GenAI, in part because of the add-ons available to subscribers. Code Interpreter, now Advanced Data Analysis, is capable of writing code, analysing data and completing complex analytical tasks. The Plug-in store offers a vast array of apps to use with GPT4, offering integration and easy access to selected favourites. GPT4 has also added a personalisation option, so that users writing long and complex prompts can save their prompt in their profile for future use, eliminating the need to compose from scratch each time. OpenAI clearly understands that humans are creatures of habit that prefer to save and use a few favourite apps rather than scouring databases or spending time composing and recomposing prompts.

Image-generating tools like DALL-E and Midjourney have grown in leaps and bounds since 2022 in terms of capabilities. Platforms like Runway and D-ID allow users to create an array of synthetic media, including videos, presentations, web sites and avatars. Improvements in NLP and NLU means ever more impressive examples of voice clones, text-to-voice generation and even text to spoken voices in other languages. Synthetic characters and multi-lingual capabilities like these offer exciting opportunities to cross linguistic and physical boundaries and to bring learners together in connected virtual spaces.

Finally, autonomous agents, widely predicted to be the Next Big Thing, do not require any prompt engineering at all. Instead, when assigned a problem to solve, agents create their own itemised task list and complete it without human instruction. Autonomous agents are the stuff of dreams and of nightmares, and an obvious target for careful regulation, but once that happens, agents will transform our conception of assignments and how we assess work.

At the time of writing, GenAI's capabilities include the following:
- Writing (GPT3.4/5, Bard, Bing, Perplexity, Claude, etc.)
- Language (voice cloning, translation, Meta's text-to-voice translation)
- Data analysis (GPT4's Code Interpreter, Advanced Data Analysis in Enterprise version)
- Creativity (text-to-image, -video, -web site; voice cloning; digital twins, avatars)
- Problem-solving (autonomous agents)

Learning innovation

If GenAI is pushing the boundaries of what can be imagined and achieved, it should inspire and push us to be creative as well. Some of the best known innovations are the product of combination thinking – that is, the combination of two pre-existing products into one new one. The smartphone and AirBnB are classic examples but there are many more. Innovation also does not necessarily require a spark of genius - it simply requires taking what we already have in front of us and looking at it in a new way. Learning innovation is no different.

The approach taken here is a classic in innovation – combination thinking: By taking digital education frameworks and mapping GenAI capabilities onto them, we can rethink course and programme designs. The focus is on active, social, collaborative and constructivist approaches to learning that address the critical need for AI competencies but do so as part of a holistic educational experience that is human-centred, community-oriented, and student-focused.

Generativism begins with two well known digital frameworks – the Community of Inquiry and the Conversational Framework – which provide the methodological approach we need to navigate this new terrain of designing learning in and for AI mediated environments. Based on community, collaboration, conversation and connection, they are also natural fits for GenAI's capabilities.

Community and collaboration

The Community of Inquiry (CoI) (Garrison, Andersen & Archer, 2000, 2001) is a theoretical framework originally conceived for the design of fully online learning. The model aims to create a deep and meaningful (collaborative-constructivist) learning experience, focusing on thinking and learning collaboratively, on the process of inquiry, and incorporating social and cognitive presence throughout. CoI defines three forms of presence, which together are key to the educational experience: Social Presence, Cognitive Presence and Teaching Presence:

- Social presence is the ability of participants to identify with the community (e.g., course of study), communicate purposefully in a trusting environment, and develop inter-personal relationships by way of projecting their individual personalities;
- Cognitive Presence is the extent to which learners are able to construct and confirm meaning through sustained reflection and discourse;
- Teaching Presence is the design, facilitation, and direction of cognitive and social processes for the purpose of realising personally meaningful and educationally worthwhile learning outcomes.

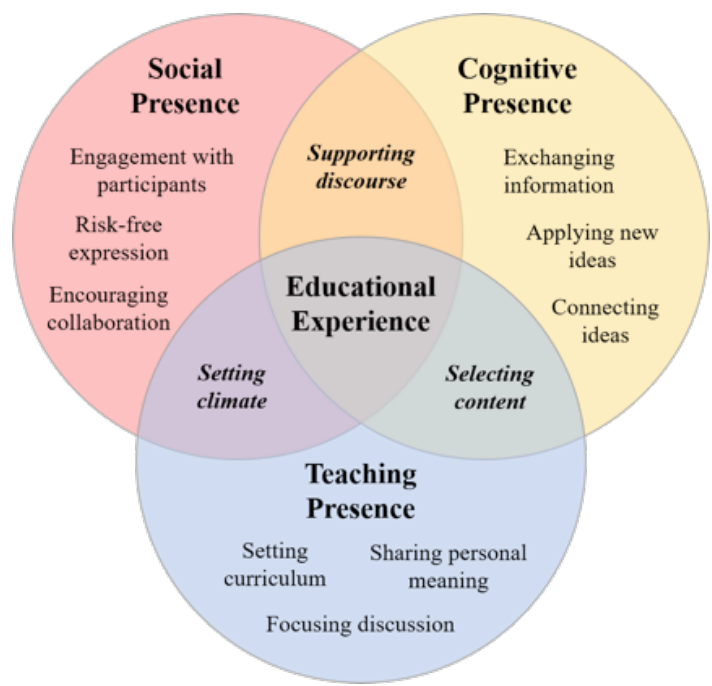

Fig 1: The Community of Inquiry Framework (Garrison, Anderson & Archer, 2000)

Generativism requires us to reimagine each of these three those forms of presence and recasts them as types of GenAI actors that interact with learners in the Community:

- Social Presence becomes Collaborator AI: students engage and work with other actors, as well as the instructor and peers.
- Cognitive Presence becomes Analytical AI: agents provide perspectives on a given topic and function as a companion, opponent and/or coach.
- Teaching Presence becomes Facilitor AI: tutors function as guides on the side that accompany and support the student throughout the course.

A CoI framework using these three GenAI actors looks like this:

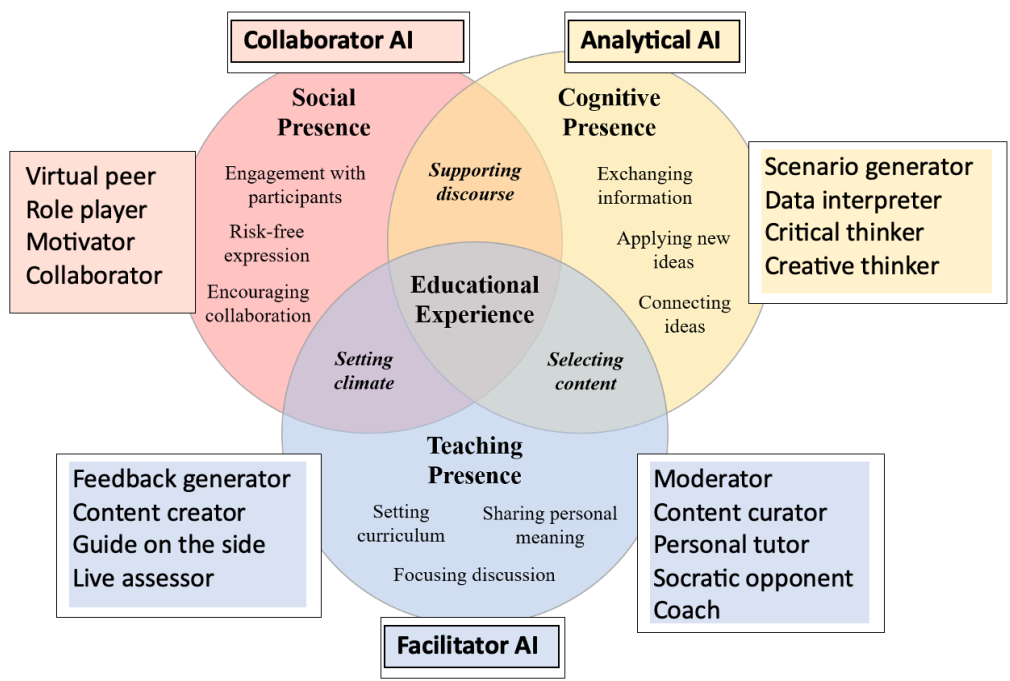

Fig 2: Community of Inquiry + GenAI

Combining the CoI framework with GenAI actors to create these new categories of Collaborative, Analytical and Facilitator AI allows us to rethink how we might design a GenAI-enabled learning experience that is inherently personalised. Adding AI tutors, twins, buddies, actors and agents brings a new dimension to the CoI's collaborative and constructivist approach, as GenAI actors act as guides, peers, experts or collaborators in the learning community, or simply playing the role of a guide-on-the-side that accompanies and supports the student as they progress through the course. GenAI actors could also be live assessors, removing the need for a summative assessment and instead replacing it with continous, GenAI-enabled assessment. (See Sharples for the list of uses of AI in digital education.) GenAI therefore offers the potential to design a truly personalised and adaptive learning experience within the established Community of Inquiry.

Conversations and constructivism

Laudrillard's Conversational Framework (CF) is another useful place to start the project of codesign with GenAI, this time focusing on the learning activities that make up each unit of learning. Laudrillard reduced the four key elements of instructionism, social learning, constructionism, collaborative learning to one model that presents the learning as taking place through reflection, feedback and clarification loops. Similar to the CoI, learning is a social and iterative process that emphasises the role of interaction and collaboration.

Laudrillard's framework was the basis for UCL's ABC Learning Design Framework, created by Natasa Petrovic and Clive Young, in 2015. ABC was orginally conceived to help educators rapidly convert face-to-face courses into an digital format but again, lends itself to any learning environment that uses the ecosystem of digital/GenAI tools. ABC divides learning activities into around six (6) types: Acquisition; Collaboration; Discussion; Investigation; Practice; Production. These original learning types are again useful starting points from which to approach designing learning activities with GenAI:

| ABC Learning Activity Type | Original Definition | + GenAI |
| --- | --- | --- |
| Acquisition | Learning through acquisition is what learners are doing when they are listening to a lecture or a podcast, reading from books or websites. | Research tools + data analysis with chatbots, private LLMs, plug-ins |
| Collaboration | Learning through collaboration embraces main discussion, practice and production. Building on investigations and acquisition, it is about taking part in the process of knowledge building itself. | Collaboration with AI tutors, buddies + guides |
| Discussion | Learning through discussion requires the learner to articulate their ideas and questions, and to challenge and respond to the ideas from the teacher, and/or from their peers. | Discussion with AI characters, experts + moderators |
| Investigation | Learning through investigation guides the learner to explore, compare and critique the texts, documents and resources that reflect the concepts and ideas being taught. | Investigation using chatbot output + student critique + data analysis |
| Practice | Learning through practice enables the learner to adapt their actions to the task goal, and use the feedback to improve their next action. Feedback may come from self-reflection, from peers, from the teacher or from the activity itself, if it shows them how to improve the result of their action in relation to the goal. | Writing tools and/or chatbot output + personal feedback, testing + iterating |
| Production | Learning through production is the way the teacher motivates the learner to consolidate what they have learned by articulating their current conceptual understanding and how they used it in practice. | Generated media and resources + iterative prompt engineering as learning. |

Here is a visual that maps GenAI tools onto each of the six ABC learning activity types:

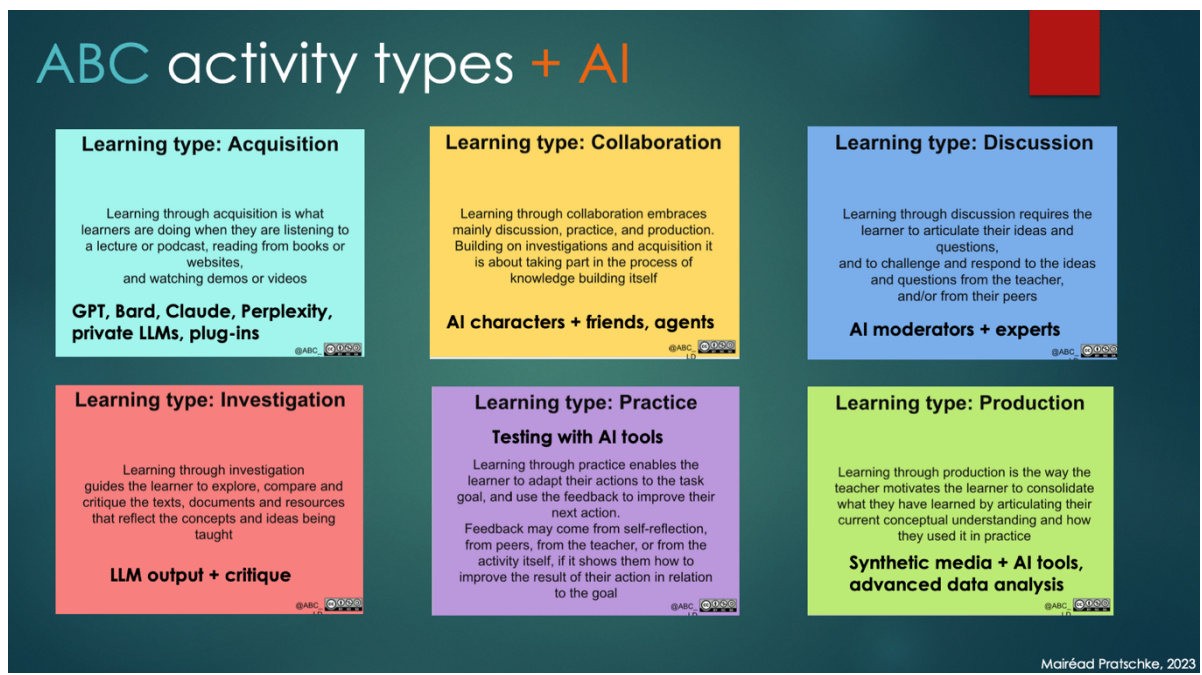

Here are those same ABC activity types + GenAI mapped onto the CoI:

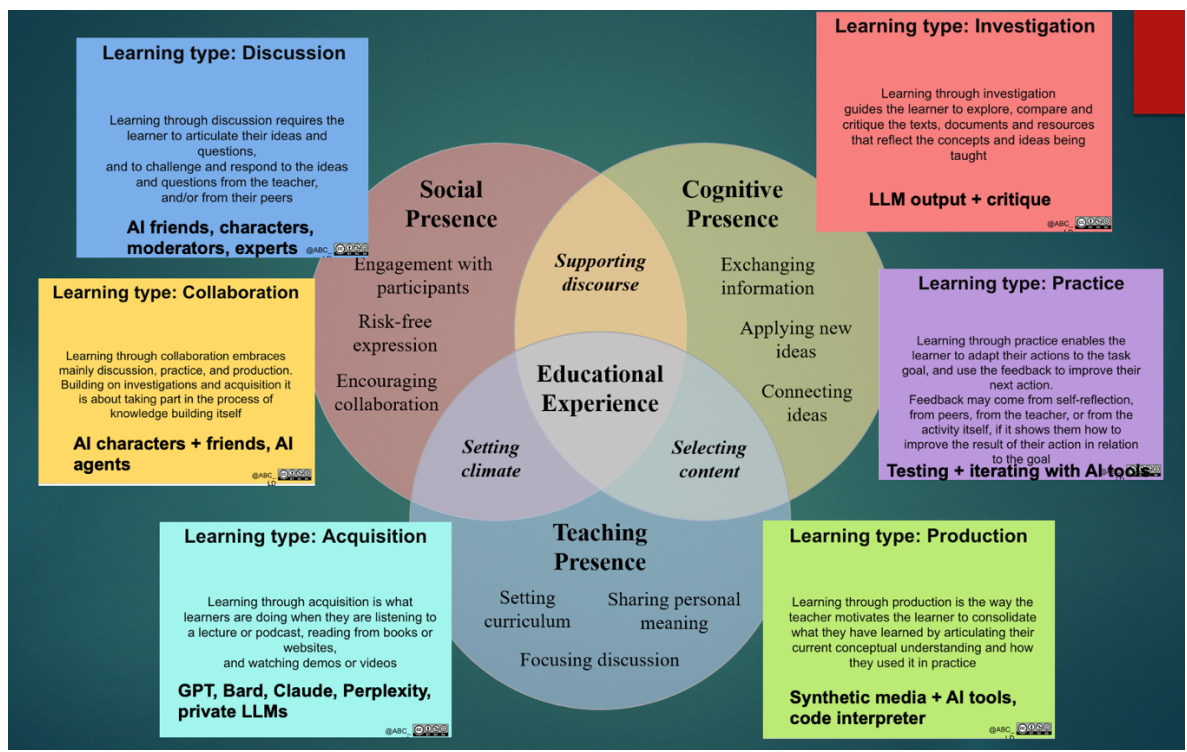

Finally, putting it all together on one diagram, CoI + ABC + GenAI:

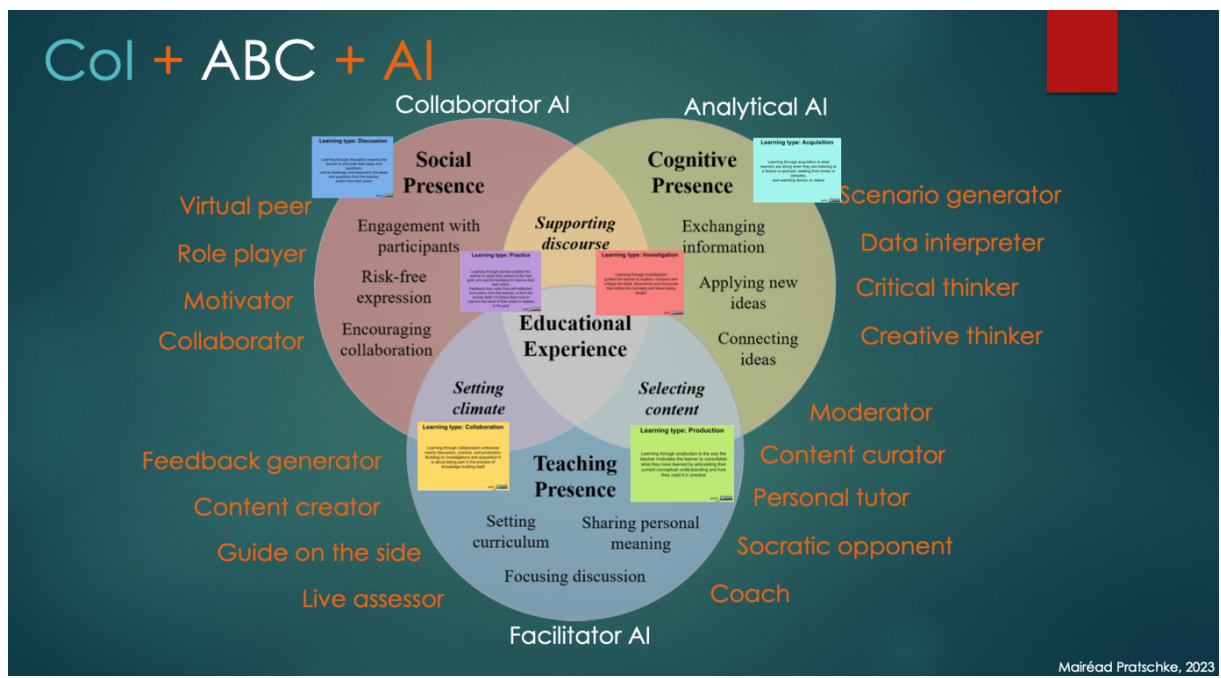

Final comments

This paper was an introduction to our emerging AI pedagogy, Generativism, a collaborative (human + GenAI) approach to learning and assessment design. Education 4.0 does not mean using technology to reinforce outmoded ways of teaching and learning. The innovation we have seen in digital education over the last two decades risks being undermined if we use GenAI to revert to older models of delivery, as we saw happen during the Covid pandemic with the use of emergency remote online teaching. GenAI, rather than increasing automation in education, should enable more interaction, more active learning, more personalisation. The focus for educators therefore needs to be on the design of active, collaborative and constructivist learning rather than the generation of content. Generativism offers an approach to designing and delivering learning experiences that are social, collaborative, community-oriented and human-centred, in collaboration with GenAI.

All comments, suggestions and feedback are most welcome, please feel free to email me at mairead.pratschke@manchester.ac.uk

References:

Hardman, Philippa (2023) The Learning Science Newletter on Substack
European Digital Education Hub (2023), Artificial Intelligence Briefing reports no. 1-4, European Union
Garrison, D. R., Anderson, T., & Archer, W. (2000). "Critical inquiry in a text-based environment Computer conferencing in higher education." *The Internet and Higher Education*, 2 (2-3), 87-105.
Garrison, D. R., Anderson, T., & Archer, W. (2001). "Critical thinking and computer conferencing: a model and tool to assess cognitive presence." *American Journal of Distance Education*, 15 (1), 7-23.
Laudrillard, Diane. (2002). *Rethinking university teaching: a conversational framework for the effective use of learning technologies (2nd ed.).* London: RoutledgeFalmer.
Laudrillard, Diane. (2012). *Teaching as a design science: building pedagogical patterns for learning and technology.* London: Routledge
Miao, Fengchun & Holmes, Wayne for UNESCO (2023), Guidance for generative AI in education and research, September 2023 ISBN 978-92-3-100612-8
Mollick, Ethan (2023) One Useful Thing newsletter on Substack
Pratschke, B. Mairéad (2023) "Blended/Hybrid Learning: The Long View"
Sharples, Mike (2023) Generative AI and Large Language Models in Digital Education, presentation at EDDTU-EU Summit, 24 May 2023

Frameworks and models
[Bloom's Taxonomy Revisited](#) (2023) Oregon State Ecampus
[Community of Inquiry](#) theoretical framework
[ABC Learning Design](#) framework

Databases
[The Rundown AI](#) tools database
[Futurepedia](#) AI tools directory